\documentclass{article}

     \usepackage{arxiv}

\usepackage[utf8]{inputenc} 
\usepackage[T1]{fontenc}    
\usepackage{hyperref}       
\usepackage{url}            
\usepackage{booktabs}       
\usepackage{amsfonts}       
\usepackage{nicefrac}       
\usepackage{microtype}      
\usepackage{graphicx}
\usepackage{float}
\usepackage{indentfirst}

\title{Kidney Recognition in CT Using YOLOv3}

\author{%
  Andréanne Lemay\\
  Polytechnique Montréal\\
  Montreal, QC H3T 1J4 \\
  \texttt{andreanne.lemay@polymtl.ca} \\
}

\begin{document}

\maketitle

\begin{abstract}
Organ localization can be challenging considering the heterogeneity of medical images and the biological diversity from one individual to another. The contribution of this paper is to overview the performance of the object detection model, YOLOv3, on kidney localization in 2D and in 3D from CT scans. The model obtained a 0.851 Dice score in 2D and 0.742 in 3D. The SSD, a similar state-of-the-art object detection model, showed similar scores on the test set. YOLOv3 and SSD demonstrated the ability to detect kidneys on a wide variety of CT scans including patients suffering from different renal conditions.
\end{abstract}

\section{Introduction}
\begin{figure}[H]

  \centering
  \graphicspath{ {./} }
  \includegraphics[scale=0.6]{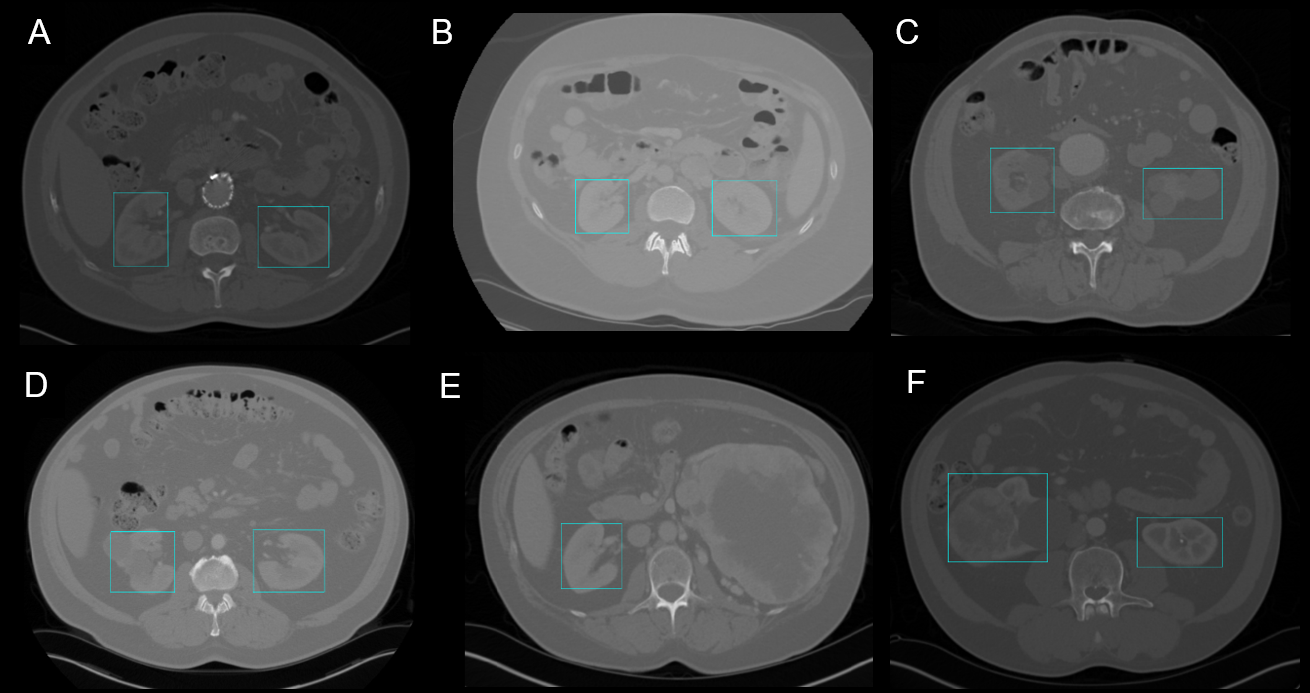}
  \caption{2D kidney detection by YOLOv3. A-B: Normal kidneys with different CT scan mean intensities. C-D: Cystic kidneys. E: Failed detection of hypertrophied kidney. F: Tumoral kidney.}
  \label{fig:ct}
\end{figure}

Organ detection is useful for various medical applications, whether it is to plan surgeries or to find pathologies. Adding bounding boxes to organs can also be the first step before applying other image processing methods like segmentation \cite{Afshari2018}, \cite{Laura2019}. Real-time organ tracking can be profitable for adaptive radiotherapy \cite{Menten2018} or laparoscopic surgeries \cite{Collins2016}. Object detection models could help with these tasks. This article focuses on 2D and 3D kidney detection. Recognition of kidneys can be challenging considering the variety of forms, textures, positionings and contrasts found in CT scans (Figure \ref{fig:ct}). 

Alongside with SSD (Single Shot Detector) and Faster R-CNN, YOLO (You Only Look Once) has proven to be a state-of-the-art and robust object detection system \cite{Laura2019}, \cite{Zhao2019}. SSD and YOLO have the advantage to be real-time models compared to the Faster R-CNN \cite{Zhao2019}. YOLO begins to appear in the medical field. The usage of YOLO \cite{Redmon} based models was recently explored for localization of normal active organs in 3D PET scans \cite{Afshari2018}, for lung nodules detection for lung cancer prevention \cite{Ramachandran2018} and for automatic nasal cavities in CT scans \cite{Laura2019}.  Considering its robustness, its speed and its accuracy on other medical images, YOLO was retained as kidney detection model. The performance of a similar object detection model, the SSD, was also quantified for this task.

\section{Methods}

YOLO takes as input 2D images which is the first challenge with its adaptation to 3D medical images. Every CT scan slice is used as a single image for inference and training. Once the bounding boxes are found on every 2D image, a 3D generalization of the non-maximum suppression algorithm was performed. This post-processing step groups the 2D boxes with a threshold criteria corresponding to their intersection over union (IoU) to generate 3D bounding boxes.

The model was trained on 14 CT scans, including 2911 512x512 2D images where 1200 contained kidneys. 41 CT scans were used to test the model (7451 images). The data come from the public KiTS2019 dataset composed of healthy and tumoral kidneys and from another dataset containing normal and cystic kidneys. The code from the following github, \footnote{https://github.com/qqwweee/keras-yolo3}, was adapted and implemented in the image processing software Dragonfly from ORS \footnote{https://www.theobjects.com/dragonfly/}. Histogram equalization was applied on every slice to increase the contrast. This step improves the detection for a broad spectrum of CT intensities.

The results are compared with the SSD using the MobileNet architecture for feature extraction. Like YOLOv3, SSD has a one-step framework \cite{Zhao2019}. The two models have the same accuracy, but YOLOv3 is about three times faster \cite{Redmon}. The Tensorflow \cite{tensorflow2015} Objection Detection API was used with Google Colaboratory to train and apply the model.

\section{Results}
\subsection{2D Detection}
YOLO showed high detection scores for the training and test sets in 2D (Dice score > 0.85) as seen in Table 1. Cystic, cancerous and healthy kidneys were recognized and located properly in CT scans largely varying in contrasts and intensities (Figure \ref{fig:ct}). The presence of artifacts caused by metal stents did not affect the performance of the model.

\textbf{Difficulty with unknown ratios.} The drop in performance from the training to the test set is mainly due to the failed recognition of kidney morphology unknown to the model. For instance, in  Figure \ref{fig:ct}E, the kidney presented on the right has an unusual size and shape. The model only recognizes and locates the other kidney. In fact, it is known that YOLO struggles to generalize to new or unknown ratio configurations \cite{Zhao2019}. The model did well on regular looking kidneys even with tumors or cysts (Figure \ref{fig:ct}C, \ref{fig:ct}D and \ref{fig:ct}F) as long as the general morphology is preserved. 

\textbf{Coarseness of bounding boxes.} Another factor affecting the scores is the coarseness of the boxes generated. When an object is localized, the box might not be properly framed or fitted (Figure \ref{fig:ct}F). It is reported that YOLOv3 struggles with perfectly aligning boxes with the detected objects \cite{Redmon}.

\textbf{Inaccuracy with kidney extremities}. Detection of the kidney extremities is challenging for the model possibly because of its smaller size or of the decrease in characteristic features. These parts are more likely to fall under the radar as seen in Figure \ref{fig:ct3d}C where the lower section of the kidneys was not detected.

The SSD and YOLOv3 have similar performances for 2D detection on the test set (Table \ref{tab:scores}). YOLOv3 has the speed advantage, taking only third of SSD’s time for inference. For an image of approximately 320 x 320 pixels, SSD takes 60 ms while YOLOv3 takes only 22 ms \cite{Redmon}. For non-time sensitive applications, SSD performs similarly to YOLOv3 (Dice: 0.851 vs 0.855 and IoU: 0.759 vs 0.747).  

\begin{table} [H]
  \caption{YOLOv3 scores for 2D and 3D kidney detection}
  \label{sample-table}
  \centering
  \begin{tabular}{lllllll}
    \toprule
    \multicolumn{1}{c}{}& \multicolumn{4}{c}{YOLOv3}& \multicolumn{2}{c}{SSD MobileNet} \\
    \cmidrule(r){2-5}
    \cmidrule(r){6-7}
    
    \multicolumn{1}{c}{}& \multicolumn{2}{c}{2D}&
    \multicolumn{2}{c}{3D} & \multicolumn{2}{c}{2D}\\
    \cmidrule(r){2-3}
    \cmidrule(r){4-5}
    \cmidrule(r){6-7}

     & Dice  & IoU  & Dice  & IoU & Dice  & IoU \\
    \midrule
    Training set (n=14) & 0.928  & 0.866  & 0.820  & 0.698 &\textbf{0.941} &  \textbf{0.889}\\
    Test set (n=41)     & 0.851  & \textbf{0.759}  & 0.742  & 0.606 & \textbf{0.855}  & 0.747\\
    \bottomrule
    \label{tab:scores}
  \end{tabular}
\end{table}

\subsection{3D Detection}
\begin{figure}[H]
  \centering
  \graphicspath{ {./} }
  \includegraphics[width=\textwidth]{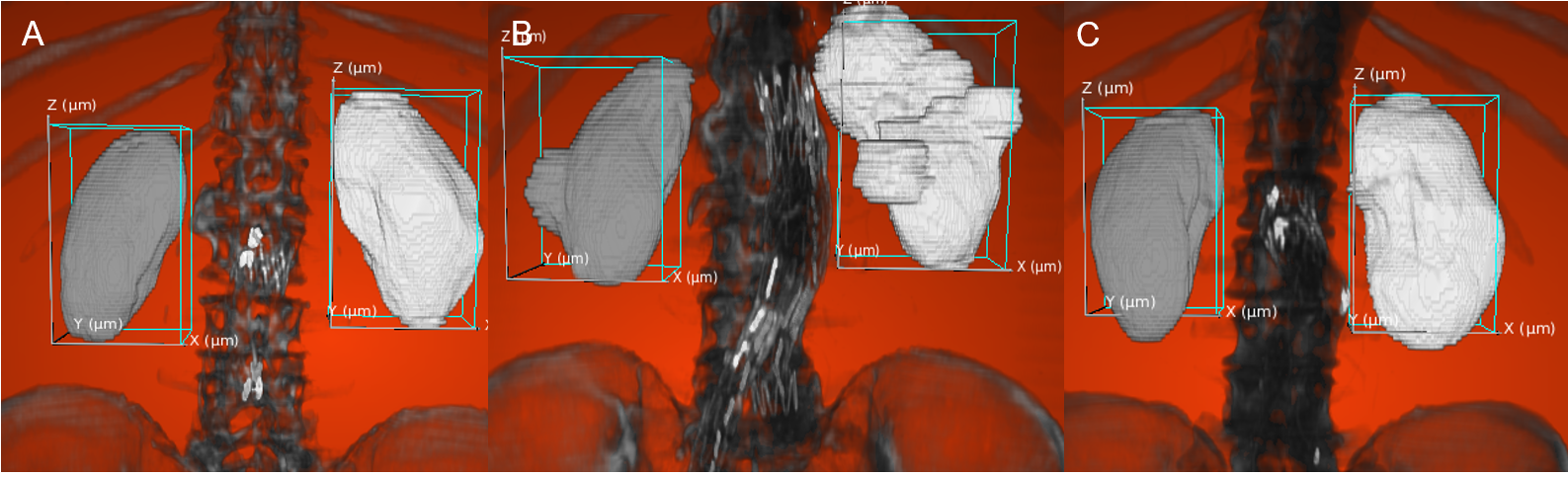}
  \caption{3D kidney detection of normal and cystic kidneys. A: Bounding box properly frames the kidneys. B-C: Overflow by the cystic (B) and healthy kidneys (C).}
\label{fig:ct3d}
\end{figure}

As seen in Figure \ref{fig:ct3d}, the 3D boxing is successful, but coarse. The boxes frame the majority of the kidney, but does not fit the organ perfectly. When transferring the results from 2D to 3D, the algorithm usually fails to contain the whole organ as shown in Figure \ref{fig:ct3d}B.

\textbf{Post-processing step unsuited for objects misaligned with axes.} One element causing this imprecision is the poor performance of the non-maximum suppression algorithm with non-axis aligned objects. When the kidney is misaligned with the Z axis, the organ overflows the bounding box.

 \textbf{Propagation of error.} Another factor leading to reduced performance in 3D detection is the propagation of error from 2D to 3D. If the 2D boxes are not representative of the kidneys, the error will propagate to the 3D boxes. This leads to lower similarity scores for 3D detection.

\section{Discussion}
Real-time object detection models like YOLOv3 or SSD showed promising results for kidney detection, suggesting that it could do well in organ recognition even with few CT scans as training set. The results demonstrate the capacity of the model to generalize to a broad range of kidney morphology. For real-time applications, YOLOv3 is more appropriate than SSD considering that, on larger images, SSD loses its real-time advantage. YOLO did show some limitations. Firstly, it is a detection model for 2D images.  Generalizing the 2D detection in 3D adds error to the kidney localization. Also, heavily cystic or pathological kidney slices were not classified properly. Finally, the boxes generated are not perfectly aligned with the objects of interest.

YOLO is a fast and robust object detection model that can also benefit the medical field for applications not requiring fine organ localization. These tasks could include detection of other organs or tumors from different modalities or real-time tracking of anatomical structures during medical procedures like radiotherapy or laparoscopic surgeries.
\newpage

\section*{Acknowledgments}
Thank you to Joseph Paul Cohen for his helpful suggestions and revisions.

\bibliographystyle{plain}
\bibliography{yolo}
 




\end{document}